# Seyfert 2 galaxies with unusually wide nebular lines

**H Winkler, T Chauke**

Dept. Physics, University of Johannesburg, PO Box 524, Auckland Park, 2006

hwinkler@uj.ac.za

**Abstract**. We report on a set of AGN that match the Seyfert 2 galaxy classification criteria, but display unusually wide "narrow" lines, with the 4959 Å and 5007 Å nebular lines overlapping with each other. This spectral line broadening is in most cases evidence of a complex profile with multiple components. It indicates an unusual narrow line region with diverse gas clouds and a range of velocity systems. We list 14 such objects with these characteristics identified in a set of Sloan Digital Sky Survey spectra. We measure the strengths of all lines visible in the spectrum, and attempt to fit multiple Gaussian profiles to the nebular lines. We quantify the line parameters of all multiple velocity systems discovered. We compare the spectral characteristics of our sample with those of other, 'conventional' type 2 Seyferts and attempt to determine whether other systematic spectral differences exist. We consider whether the investigated sample constitutes a clear sub-class of the Seyfert 2 population. In conclusion we offer possible explanations for the unusual line profiles.

## 1. Introduction

Seyfert 2 galaxies are characterised by a nuclear emission line spectrum marked by [O III] 5007 Å and 4959 Å (the so-called nebular lines) that are significantly stronger than the Hβ line, and [N II] 6583 Å of a similar strength as Hα. In contrast to the related Seyfert 1 galaxies, type 2 nuclei display no overt signs of broadened hydrogen and helium lines [1]. It has however been established that a significant fraction of type 2 Seyferts, including the 'prototype' NGC 1068, do in fact contain a broad line region that is mostly hidden behind dust [2].

The nebular lines, along with other 'narrow' lines in Seyfert spectra, form relatively far from the central black hole whose immediate surroundings generate the ionising radiation responsible for the emission lines. The dynamics of the narrow emission line region manifests itself in the line profile, and hence the study of the profiles of prominent spectral features such as the nebular lines has been a favoured method to map the outer parts of a Seyfert nucleus [3, 4, 5]. In some instances the line profile was shown to be asymmetric and broader than expected. Indeed, the nebular lines in the NGC 1068 nuclear spectrum are so wide that they partly overlap [2].

In this work we investigate a set of Seyfert 2 galaxies characterised by unusually wide nebular lines with complex profiles. We attempt to identify individual components making up these profiles and measure the relative strengths of other lines in the spectrum. We compare the properties thus determined to a sample of Seyfert 2s that do not show evidence of nebular line complexity. We probe possible systematic differences between the samples, and explore possible causes of the complex line profile. We determine whether there are sufficient grounds for assigning the objects studied to a new sub-class of Seyferts.

## 2. Spectral data reductions

2.1. Identification of sample

During the course of a programme initiated by one of us to classify AGN observed by the Sloan Digital Sky Survey (SDSS) [6], a number of spectra were identified that matched the typical Seyfert 2 spectral line ratio criteria ([O III] 5007 Å >> H-beta, [N II] 6583 Å ~ H-alpha) [7]. Some of these spectra were however noted to be unusual, as their nebular lines exhibited a complex profile, to the extent where these lines partly overlapped. An example of the spectrum of such an object is illustrated in figure 1 below.

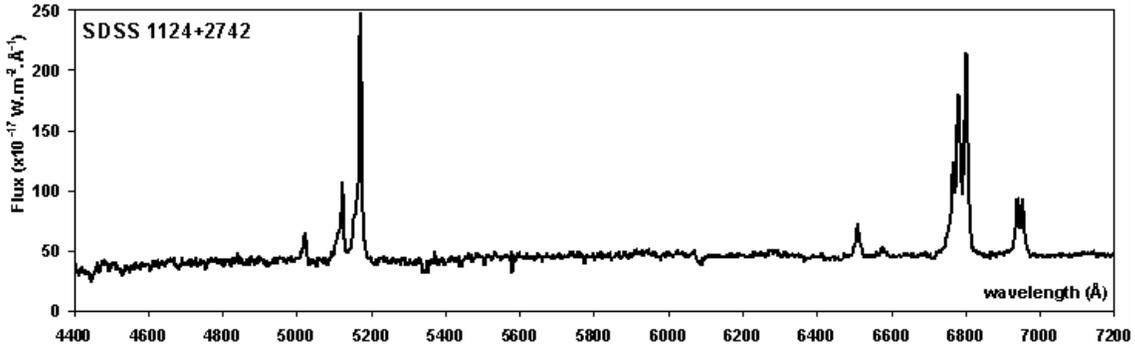

**Figure 1.** Example of a spectrum of a Seyfert 2 galaxy with wide nebular lines. Note that the 4959 Å and 5007 Å lines (redshifted into the range 5100-5200 Å in the diagram above) overlap each other.

We resolved to study this potential sub-class of Seyfert galaxies further, and identified 12 objects from the SDSS spectral collection that met the overlapping nebular line profile criterion. These are listed in table 1. We also chose five objects with regular Seyfert 2 spectra to act as a comparison sample. The details of the comparison sample are given in table 2.

**Table 1.** List of AGN of the class identified

| code | R.A.(2000)  | Dec(2000)  | $z$   |
|------|-------------|------------|-------|
| A    | 02h08m23.8s | −00 20 00  | 0.074 |
| B    | 08h11m21.4s | +40 54 51  | 0.067 |
| C    | 08h13m47.5s | +49 41 10  | 0.094 |
| D    | 09h18m07.5s | +34 39 45  | 0.097 |
| E    | 10h56m38.9s | +14 19 30  | 0.081 |
| F    | 11h24m23.9s | +27 42 45  | 0.032 |
| G    | 11h47m19.9s | +07 52 43  | 0.083 |
| H    | 12h17m27.8s | +15 54 13  | 0.084 |
| I    | 12h17m41.9s | +03 46 31  | 0.080 |
| J    | 14h48m38.5s | +10 55 36  | 0.089 |
| K    | 14h50m18.7s | +12 06 46  | 0.098 |
| L    | 15h11m41.3s | +05 18 10  | 0.084 |

**Table 2.** Comparison sample

| code | R.A.(2000)  | Dec(2000)  | $z$   |
|------|-------------|------------|-------|
| M    | 00h17m41.8s | +00 07 53  | 0.070 |
| N    | 02h02m23.7s | +12 47 17  | 0.086 |
| O    | 07h37m15.8s | +31 31 11  | 0.027 |
| P    | 10h22m12.6s | +38 37 43  | 0.056 |
| Q    | 14h24m05.5s | +01 47 57  | 0.056 |

## 2.2. Spectral reductions

Electronic wavelength- and flux-calibrated spectra were downloaded from the SDSS website. The spectra were then converted to the target's rest frame by eliminating the redshift. High resolution representations of the spectral region spanning the 4959 Å, 5007 Å nebular lines are shown in figure 2.

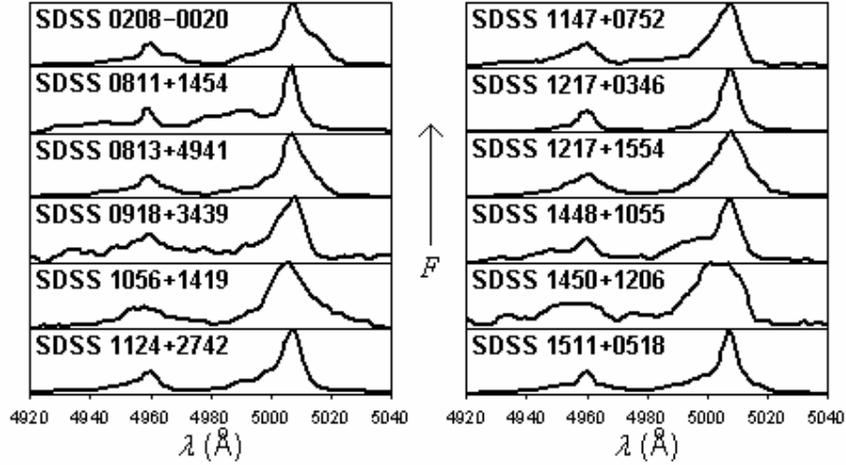

**Figure 2.** The profiles of the [O III] 4959 Å, 5007 Å lines. The vertical scale in each case ranges from the lowest to the highest values in the displayed wavelength window.

## 2.3. Spectral line strengths

To systematically compare the relative strengths of the emission lines in each spectrum, we adopted a new procedure to parameterise SDSS AGN spectra. In essence, this procedure measures $\Phi(\lambda)$, the integrated flux for each line over the wavelength range corresponding to a velocity offset $|\Delta v| < 380$ km/s relative to the galaxy's recession velocity (corresponding to 5 SDSS wavelength bins on either side of the peak).

## 3. Spectral line analysis

### 3.1. Asymmetry index

The asymmetry of a spectral line may be quantified according to the procedures of Heckman et al [3] or Whittle [4]. The latter method is based on the area under a curve, and is thus less appropriate when spectral lines overlap. We hence preferred the Heckman et al procedure, as it only requires the identification of the wavelengths where the spectral lines reach 20% ($\lambda_{B,20}$ on the blue side, $\lambda_{R,20}$ on the red side) and 80% of its maximum. $\lambda_{C,80}$ is the midpoint between $\lambda_{B,80}$ and $\lambda_{R,20}$ (see figure 3).

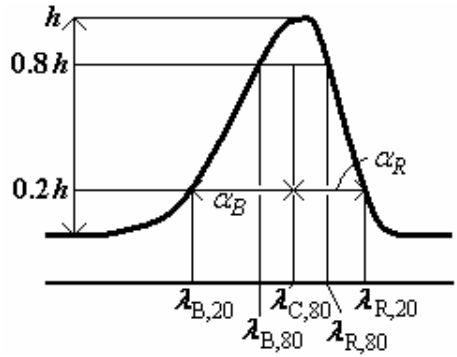

**Figure 3.** Diagram to illustrate the definition of the Heckman et al [3] asymmetry index.

The Asymmetry index at 20% peak height is then defined by

$$AI_{20} = \frac{\alpha_B - \alpha_R}{\alpha_B + \alpha_R} = \frac{(\lambda_{C,80} - \lambda_{B,20}) - (\lambda_{R,20} - \lambda_{C,80})}{(\lambda_{C,80} - \lambda_{B,20}) + (\lambda_{R,20} - \lambda_{C,80})} = \frac{\lambda_{B,80} + \lambda_{R,80} - \lambda_{B,20} - \lambda_{R,20}}{\lambda_{R,20} - \lambda_{B,20}}.$$

This asymmetry index is listed for all objects studied in table 3.

**Table 3.** Measured parameters for the Seyfert 2 galaxies studied here.

| code | $AI_{20}$ | 1st cmp. FWHM | 2nd cmp. offset | 2nd cmp. FWHM | $\rho_{3727}$ | $\rho_{3868}$ | $\rho_{4686}$ | $\rho_{4861}$ | $\rho_{6583}$ | $\rho_{6300}$ | $\rho_{6725}$ |
|---|---|---|---|---|---|---|---|---|---|---|---|
| A | 0.02 | 380 | 360 | 1400 | 1.28 | 0.97 | - | 0.99 | 1.61 | 0.77 | 1.41 |
| B | - | 320 | 930 | 1800 | 1.90 | 1.22 | - | - | 1.81 | 1.23 | 1.89 |
| C | 0.13 | 350 | 60 | 1200 | 1.41 | 0.83 | 0.33 | - | - | 1.10 | 1.53 |
| D | 0.89 | 300 | 210 | 800 | 1.85 | 1.38 | 1.12 | - | 2.83 | 1.34 | 2.05 |
| E | −0.45 | 600 | −180 | 1500 | 1.96 | 0.92 | - | - | 2.32 | 1.65 | 2.32 |
| F | 1.75 | 350 | 270 | 1200 | 1.50 | 0.94 | 0.26 | 1.18 | 2.03 | 1.22 | 1.78 |
| G | 0.69 | 330 | 270 | 800 | 1.55 | 1.15 | 0.61 | 1.52 | 2.08 | 1.02 | 1.66 |
| H | 0.35 | 330 | 140 | 1100 | 1.62 | 1.13 | 0.44 | 0.84 | 1.92 | 0.97 | 1.62 |
| I | 0.59 | 510 | 120 | 1400 | 1.23 | 1.01 | 0.45 | 0.70 | 1.93 | 0.89 | 1.39 |
| J | 2.95 | 390 | 630 | 1200 | 1.76 | 1.16 | - | - | 2.52 | 1.26 | 1.89 |
| K | 0.47 | 400 | 570 | 900 | 2.13 | 1.16 | - | - | 2.71 | 1.53 | 2.11 |
| L | 0.39 | 310 | 250 | 1500 | 1.38 | 1.02 | 0.66 | 1.72 | - | 0.97 | 1.50 |
| M | −0.34 | 340 | - | - | 1.36 | 0.93 | 0.40 | 0.80 | 1.90 | 0.87 | 1.43 |
| N | −0.28 | 350 | - | - | 1.05 | 0.87 | 0.52 | 1.05 | 1.45 | 0.43 | 1.10 |
| O | −0.28 | 330 | - | - | 1.12 | 0.88 | 0.34 | 0.73 | 1.64 | 0.69 | 1.35 |
| P | 0.26 | 220 | - | - | 1.45 | 0.95 | - | - | 1.28 | 0.75 | 1.22 |
| Q | 1.29 | 300 | - | - | 1.49 | 0.96 | 0.60 | −0.07 | 1.68 | 0.74 | 1.48 |

3.2. Line positions and widths

We fitted multiple Gaussian profiles to the nebular lines and to the region around Hα. These take the mathematical form

$$F(\lambda) = F(\lambda_0) \exp\left(-\frac{(\lambda - \lambda_0)^2}{2\sigma^2}\right)$$

where $\lambda_0$ is the peak wavelength, $F(\lambda_0)$ is the peak flux and the full width at half-maximum (FWHM) corresponds to $2.355\sigma$.

We first fitted Gaussians with widths corresponding to a common velocity and peak wavelengths corresponding to identical recession velocities to the [O III], [N II] and Hα lines. We constrained the peak heights of the two nebular lines, and of the two [N II] lines, to stick to the theoretical 3:1 ratio. We then fitted a second component (of corresponding widths and peak ratio) to the nebular lines. We also found that we required a further, broader component at Hα to obtain a reasonable match with the observed profile. Details of the parameters obtained through these fits are also given in table 3, and an example of the fitted profiles may be viewed in figure 4.

3.3. Spectral parameters

Table 3 also lists the ratio of diagnostically significant [7] other spectral lines ([O II] 3727 Å, [Ne III] 3868 Å, He II 4686 Å, H I 4861 Å, [N II] 6583 Å, [O I] 6300 Å and the [S II] 6725 Å doublet) relative to the stronger 5007 Å nebular line. These are represented in terms of the parameter $\rho_\lambda = 2 + \log[\Phi(\lambda)/\Phi(5007)]$. Note that for the [S II] doublet we use $\Phi(6725) = \Phi(6716) + \Phi(6731)$.

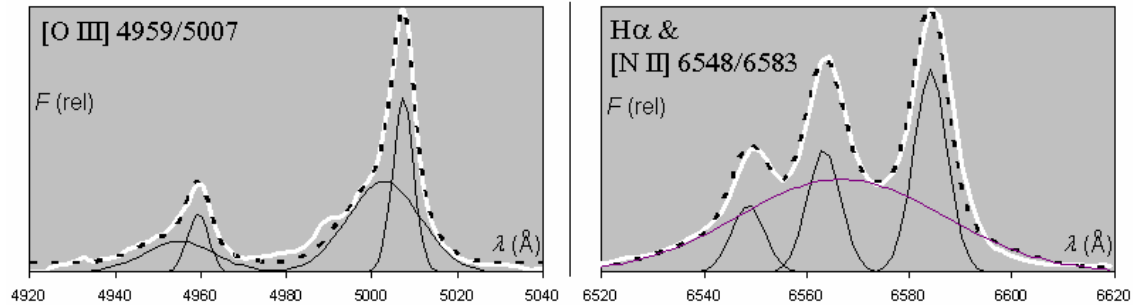

**Figure 4.** Gaussian components for the nebular lines ([O III] 5007 Å and 4959 Å), Hα and the nitrogen lines [N II] 6583 Å and 6548 Å (black solid line), for the galaxy SDSS 1124+2742. The dashed line represents the sum of the Gaussian components, while the white line is the actual data. The vertical range and scale are adjusted to span the lowest value and highest peak in the wavelength range viewed.

## 4. Discussion

4.1. Spectral comparison with other Seyfert 2 galaxies
To evaluate whether significant spectral parameter differences exist between the sample of objects with broadened nebular lines (codes A-L in table 3) and the control sample (codes M-Q), we performed Student's t-tests with the values for the asymmetry parameters, widths and line ratios.

At the 95% confidence level, we find significant differences between the samples for the line ratio parameters $\rho_{3727}$, $\rho_{3868}$, $\rho_{6300}$ and $\rho_{6725}$, but not for the widths, the asymmetry indexes or $\rho_{4686}$. At the same time, we note a wide spread in all parameters for the objects coded A-L, whereas there is more uniformity in the control sample.

4.2. A new sub-class of Seyfert galaxies?
The findings of this study confirm that the Seyfert 2 galaxies with broad nebular lines do not constitute a homogeneous group. Some of the objects studied are definitely unusual, and warrant further investigation. However, unless similar features are detected in several other objects, these cannot be defined as a new Seyfert sub-class.

4.3. Explaining the complex profiles
In most cases the asymmetries and second components identified point to at least parts of the emission line regions being blue-shifted compared to the system velocity. This signifies gas moving towards Earth. We also note that a broad-line component in Hα was fitted in all objects investigated. Although this component was usually too weak to have been evident in a low-resolution view of the spectrum, it does support the view that type 2 Seyferts are nothing but broad-line objects with a highly obscured nucleus, meaning that the broad component is generally too faint to be detected.

It is thus plausible for many of the objects in our sample the Earthwards moving gas is part of a jet, and that the outflow in the opposite direction is not visible due to obscuration by dust. We may also draw on previous high angular resolution studies of the closest and brightest AGN known with overlapping nebular lines, NGC 1068 [8], and another well-known Seyfert 2, NGC 5643 [9]. These studies reveal clumps of matter that to some extent suggest a bi-conical structure with matter outflow, though other mechanisms cannot be entirely excluded.

Alternatively, another possible explanation is that the SDSS spectra have combined the spectra of two interacting galaxies with emission lines, possibly even Seyfert nucleus pairs [10].

In either event, the objects studied here all display a complex narrow-line region structure, and the more abnormal ones would make good candidates for more in-depth study using high-resolution imaging, polarimetry or radio observations.

**Acknowledgements**
This paper utilized data from the Sloan Digital Sky Survey (SDSS). Funding for the SDSS and SDSS-II has been provided by the Alfred P. Sloan Foundation, the Participating Institutions, the National Science Foundation, the U.S. Department of Energy, the National Aeronautics and Space Administration, the Japanese Monbukagakusho, the Max Planck Society, and the Higher Education Funding Council for England. The SDSS Web Site is http://www.sdss.org/.

The SDSS is managed by the Astrophysical Research Consortium for the Participating Institutions. The Participating Institutions are the American Museum of Natural History, Astrophysical Institute Potsdam, University of Basel, University of Cambridge, Case Western Reserve University, University of Chicago, Drexel University, Fermilab, the Institute for Advanced Study, the Japan Participation Group, Johns Hopkins University, the Joint Institute for Nuclear Astrophysics, the Kavli Institute for Particle Astrophysics and Cosmology, the Korean Scientist Group, the Chinese Academy of Sciences (LAMOST), Los Alamos National Laboratory, the Max-Planck-Institute for Astronomy (MPIA), the Max-Planck-Institute for Astrophysics (MPA), New Mexico State University, Ohio State University, University of Pittsburgh, University of Portsmouth, Princeton University, the United States Naval Observatory, and the University of Washington.